\documentclass[a4paper,11pt]{article}
\usepackage{pos}

\usepackage[utf8]{inputenc}
\usepackage{amsmath}
\usepackage{amsfonts}
\usepackage{graphicx}
\usepackage{hyperref}
\usepackage{braket}
\usepackage{multirow}
\usepackage{enumitem}
\usepackage{bbm}
\usepackage{units}
\usepackage{tikz}
\usepackage{subfloat}
\usetikzlibrary{fadings,shadows,positioning,calc,matrix,shapes,decorations.markings,shapes.arrows}

\graphicspath{{plots/}}

\definecolor{red}{rgb}{.9,0,0}
\definecolor{blue}{rgb}{0,0,.75}
\definecolor{green}{rgb}{0,.6,0}
\definecolor{magenta}{cmyk}{0,1,0,0}
\definecolor{myyellow}{rgb}{.98,.84,.37}

\newcommand{\red}{\color{red}}
\newcommand{\blue}{\color{blue}}

\title{Tensor-network study of the 3d $O(2)$ model at non-zero chemical potential and temperature}
\ShortTitle{Tensor-network study of the 3d $O(2)$ model at non-zero $\mu$ and $T$}
 
\author*[a]{Jacques Bloch}
\author[a,b]{Robert Lohmayer}
\author[a]{Maximilian Meister}

\affiliation[a]{Institute for Theoretical Physics, University of Regensburg, Regensburg, Germany}

\affiliation[b]{RCI Regensburg Center for Interventional Immunology, Regensburg, Germany}

\emailAdd{jacques.bloch@ur.de}
\emailAdd{robert.lohmayer@ur.de}
\emailAdd{maximilian.meister@ur.de}

\abstract{We present results of tensor-network simulations of the three-dimensional $O(2)$ model at non-zero chemical potential and temperature, which were computed using the higher-order tensor-renormalization-group method (HOTRG). This necessitated enhancements to the HOTRG blocking procedure to reduce the truncation error in the case of anisotropic tensors. Moreover, the construction of the truncated vector spaces was adapted to strongly reduce the effect of systematic errors in the computation of observables
using the finite-difference method. Our (improved) HOTRG results for the evolution of the number density with the chemical potential are in agreement with results obtained with the worm algorithm, and both the Silver Blaze phenomenon at zero temperature and the temperature dependence of the number density can be adequately reproduced.}

\FullConference{%
 The 38th International Symposium on Lattice Field Theory, LATTICE2021
  26th-30th July, 2021
  Zoom/Gather@Massachusetts Institute of Technology
}

\begin{document}
\maketitle

\section{Introduction}

In this talk we present first tensor-network results for the three-dimensional $O(2)$ model in thermal equilibrium at non-zero chemical potential and temperature \cite{Bloch:2021mjw}. We validate our results using benchmark data obtained with the worm algorithm \cite{Prokofiev:2001zz,Banerjee:2010kc,Langfeld:2013kno}.

Simulations of classical or quantum spin systems in thermal equilibrium on a d-dimensional lattice are typically performed using importance sampling Monte Carlo (MC) methods, which are stochastic in nature.
Recently an alternative, deterministic method has been considered which reformulates the partition function $Z$ as a fully contracted tensor network. On such a tensor network, the higher-order tensor-renormalization-group method (HOTRG) is used to compute observables numerically.

An important motivation to consider tensor-network methods is the possibility to apply these methods when the action is complex, for example in the presence of a chemical potential, and MC simulations fail because of the sign problem \cite{Aarts:2013bla}.

To reformulate the partition function as a tensor network, one introduces dual variables and integrates out the original degrees of freedom. This procedure is similar to that encountered when applying the worm algorithm. However, the computation of observables is very different. Whereas the worm algorithm uses a closed-path Monte Carlo sampling, the tensor-network methods are based on approximations using singular value decompositions (SVD) and are deterministic in nature \cite{DeLathauwer2000,Levin:2006jai,Xie_2012}.
Although the application of tensor methods to statistical physics models is quite new, it has already been applied to a variety of spin models and gauge theories \cite{Akiyama:2019xzy,Yu:2013sbi,Zou:2014rha,Yang_2016,Bazavov:2015kka,Kuramashi:2018mmi,Kuramashi:2019cgs}, mainly in two dimensions but some also in three and four dimensions, see Ref.~\cite{Meurice:2020pxc} for a review. Herein we present the first tensor-network study of the three-dimensional $O(2)$ model \cite{Bloch:2021mjw}.


\section{Tensor formulation of the 3d $\boldsymbol{O(2)}$ model}

The $O(2)$-symmetric action of this three-dimensional quantum model with chemical potential is given by
\begin{align}
S = - \beta \sum_{k=1}^V \sum_{\nu=1}^3 \cos(\theta_k-\theta_{k+\hat\nu} - i \mu\delta_{\nu3}) 
\end{align}
where $k$ is a linear index defined on the cubic $N_s^2\times N_t$ lattice, $V$ is its volume, $\beta$ is the coupling,
$\mu$ the chemical potential acting in the time direction, and $\hat\nu$ a unit step in the $\nu$-direction. The partition function for the model is
\begin{align}
Z = \int_0^{2\pi}\frac{d\theta_1}{2\pi} \dots \int_0^{2\pi} \frac{d\theta_V}{2\pi} \,  e^{-S(\theta_1,\cdots,\theta_V)} .
\end{align} 
When $\mu$ is non-zero, the action is complex and MC simulations encounter a strong sign problem. 
To reformulate the partition function as a fully contracted tensor network, one factorizes the Boltzmann probability over the lattice links and rewrites the exponential factor on each link using the Jacobi-Anger expansion
\begin{align}
e^{\beta\cos\theta} = \sum_{n=-\infty}^\infty I_{n}(\beta) \, e^{i n \theta}  ,
\end{align}
where $I_{n}(\beta)$ are modified Bessel functions of the first kind.
This procedure introduces new link variables $n_{k,\nu}$ on all links connecting the sites $k$ and $k+\hat\nu$. The partition function can then be reordered such that the multidimensional spin integral factorizes in individual spin integrals 
\begin{align}
\int_0^{2\pi} \frac{d\theta}{2\pi} e^{i n \theta} = \delta_{n,0} ,
 \quad n \in \mathbb{Z} .
\end{align}
Once the spins are integrated out, the partition function can be reformulated as a complete contraction of a tensor network,
\begin{align}
Z = \sum_{\{n\}} \prod_{k=1}^V T_{n_{k-\hat 1,1}n_{k,1}n_{k-\hat 2,2}n_{k,2} n_{k-\hat 3,3}n_{k,3}}
\end{align}
with local tensors of order six,
\begin{align*}
T_{n_{k-\hat 1,1}n_{k,1}n_{k-\hat 2,2}n_{k,2} n_{k-\hat 3,3}n_{k,3}} = \sqrt{ e^{\mu (n_{k,3}+ n_{k-\hat 3,3})} \prod_{\nu=1}^3 I_{n_{k,\nu}}(\beta) I_{n_{k-\hat\nu,\nu}}(\beta) } \,\,
\times \hspace{-5mm} \underbrace{\delta_{\Delta n_k,0}}_{\text{current conservation}}
\end{align*}
where $\Delta n_k$ is the discretized divergence of the $O(2)$ current,
\begin{align*}
\Delta n_k = \sum_{\nu=1}^3 (n_{k,\nu}-n_{k-\hat\nu,\nu}) .
\end{align*}

Although formally $n\in\mathbb Z$, one truncates the range of $n$ in practice by keeping only the $D$ largest weights $I_{n}(\beta)\,e^{\mu n \delta_{\nu,3}}$ in the construction of the initial local tensor, which then has total dimension $D^6$.

To compute thermodynamic observables one has to compute derivatives of $\ln Z$ with respect to the parameters of the model. In the tensor formulation one can either perform numerical derivatives using finite differences of computed values of $\ln Z$ or one can take analytical derivatives of the tensor network, which then yields a formula involving the pure tensor and one or more so-called impurity tensors. Both methods have their own problems, which are related to systematic errors in the contraction of tensor networks.

\section{Higher order tensor renormalization group}

\begin{figure}
\normalsize
\parbox{0.28\textwidth}{\centering\includegraphics[scale=0.78]{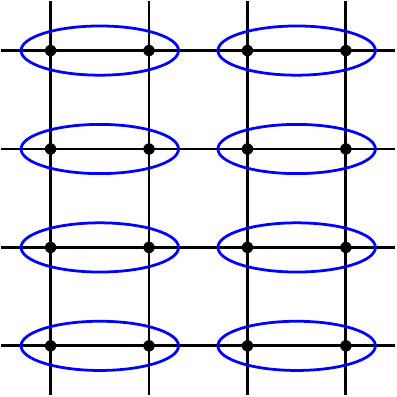}}
$\to$
\parbox{0.16\textwidth}{\centering\includegraphics[scale=0.78]{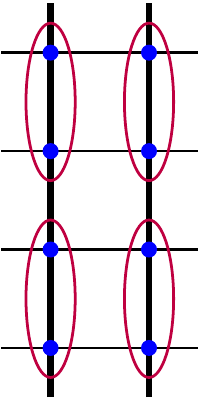}}
$\to$
\parbox{0.16\textwidth}{\centering\includegraphics[scale=0.78]{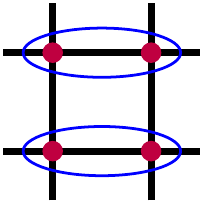}}
$\to$
\parbox{0.1\textwidth}{\centering\includegraphics[scale=0.78]{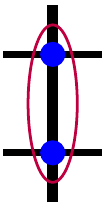}}
$\to$
\parbox{0.1\textwidth}{\centering\includegraphics[scale=0.78]{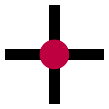}}

\scriptsize
\parbox{0.28\textwidth}{\centering{$T^{(0)}\to {\blue T^{(1)}}$}}
\normalsize\phantom{$\to$}
\scriptsize
\parbox{0.16\textwidth}{\hspace{6mm}{\blue$T^{(1)}\to {\red T^{(2)}}$}}
\normalsize\phantom{$\to$}
\scriptsize
\parbox{0.16\textwidth}{\hspace{6mm}{\red$T^{(2)}\to {\blue T^{(3)}}$}}
\normalsize\phantom{$\to$}
\scriptsize
\parbox{0.16\textwidth}{\hspace{2mm}{\blue$T^{(3)}\to {\red T^{(4)}}$}}
\normalsize\phantom{$\to$}
\scriptsize
\parbox{0.1\textwidth}{\hspace{-1mm}{\red$T^{(4)}$}}
\caption{Typical blocking procedure in 2d HOTRG using alternating contraction directions. After each blocking a new local tensor is constructed.}
\label{Fig:hotrg}
\end{figure}

In HOTRG the full tensor-network contraction is performed using an iterative blocking procedure \cite{Xie_2012}. An example of this procedure for a $4\times 4$ lattice in two dimensions is shown in Fig.\ \ref{Fig:hotrg}. In this procedure two adjacent tensors of coarsening level $i$ are contracted over their shared link in direction $\nu$ to yield a tensor of coarsening level $i+1$,
\begin{align}
T^{(i)} \odot_\nu T^{(i)} \to T^{(i+1)} .
\end{align}
When the tensor network has been blocked iteratively until only a single tensor remains, the latter is traced over its forward-backward directions to produce the partition function.

\begin{figure}
\centerline{
\hspace{5mm}\parbox{0.25\textwidth}{\centering\includegraphics[scale=1.8]{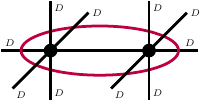}}
\hspace{8mm}$\longrightarrow$
\parbox{0.25\textwidth}{\centering\includegraphics[scale=1.8]{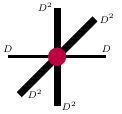}}
\hspace{-2mm}$\stackrel{U^{(y)},U^{(z)}}{\longrightarrow}$\hspace{-2mm}
\parbox{0.25\textwidth}{\centering\includegraphics[scale=1.8]{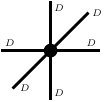}}
}\vspace{-2mm}
\centerline{
\hspace{5mm}\parbox{0.3\textwidth}{\hspace{12mm}\scriptsize $T^{(i)}$ \hspace{11mm} $T^{(i)}$}
\phantom{\hspace{8mm}$\longrightarrow$}
\parbox{0.25\textwidth}{\hspace{13mm}\scriptsize $M$}
\phantom{\hspace{-2mm}$\stackrel{U^{(y)},U^{(z)}}{\longrightarrow}$\hspace{-2mm}}
\parbox{0.25\textwidth}{\hspace{13mm}\scriptsize $T^{(i+1)}$}
}
\caption{Contraction of two local tensors in the three-dimensional case. On the left two local tensors with modes of dimension $D$ are contracted over their shared link along the $x$-direction to yield the tensor $M$ with fat modes of dimensions $D^2$ in the directions perpendicular to the contraction direction.
The frames are then applied to the fat modes to construct the new local tensor $T$ on the right, where all modes have again dimension $D$.}
\label{Fig:hotrg3d}
\label{Fig:hosvd}
\end{figure}

However, if we look at the dimensionality of the tensor contractions we notice that some additional approximations are required.
When contracting two local tensors, as shown in Fig.\ \ref{Fig:hotrg3d} for three dimensions, the modes perpendicular to the contraction direction can be gathered in fat modes and the dimensionality of the new blocked tensor grows. Two tensors $T$ of dimension $D^6$ are contracted along the $x$-direction into a tensor $M$ of dimension $D^{10}$ where the fat modes have dimension $D^2$,
\begin{align}
T^{(i)} \odot_x T^{(i)} \longrightarrow M .
\end{align}
The tensor $M$ cannot be used as such in further blockings because this would give rise to an exponential growth of the tensor dimension and of the computational cost. 
To keep the dimensionality under control, the $D^2$ dimensional fat modes of $M$ are truncated by projecting them on $D$-dimensional subspaces, such that the new local tensor again has dimension $D^6$. This is achieved using $D^2 \times D$ semi-orthogonal matrices that are constructed using the singular value information obtained from SVDs of the matrix unfoldings of $M$ with respect to all its fat modes, in analogy to the higher order singular value decomposition (HOSVD) \cite{DeLathauwer2000}.
However, a peculiarity of  the HOTRG blocking procedure is that the forward and backward fat modes must be projected on the same subspaces. 

The procedure is illustrated in Fig.\ \ref{Fig:hosvd} for a contraction in the $x$-direction where the truncation is performed using the $D^2 \times D$ matrices  $U^{(y)}$ and $U^{(z)}$, also called frames, and which yields the new local tensor
\begin{align}
M \stackrel{U^{(y)},U^{(z)}}{\longrightarrow} T^{(i+1)} .
\end{align}



\section{Results}

\subsection{Specific heat}

\begin{figure}
\centering
\includegraphics[width=0.49\textwidth]{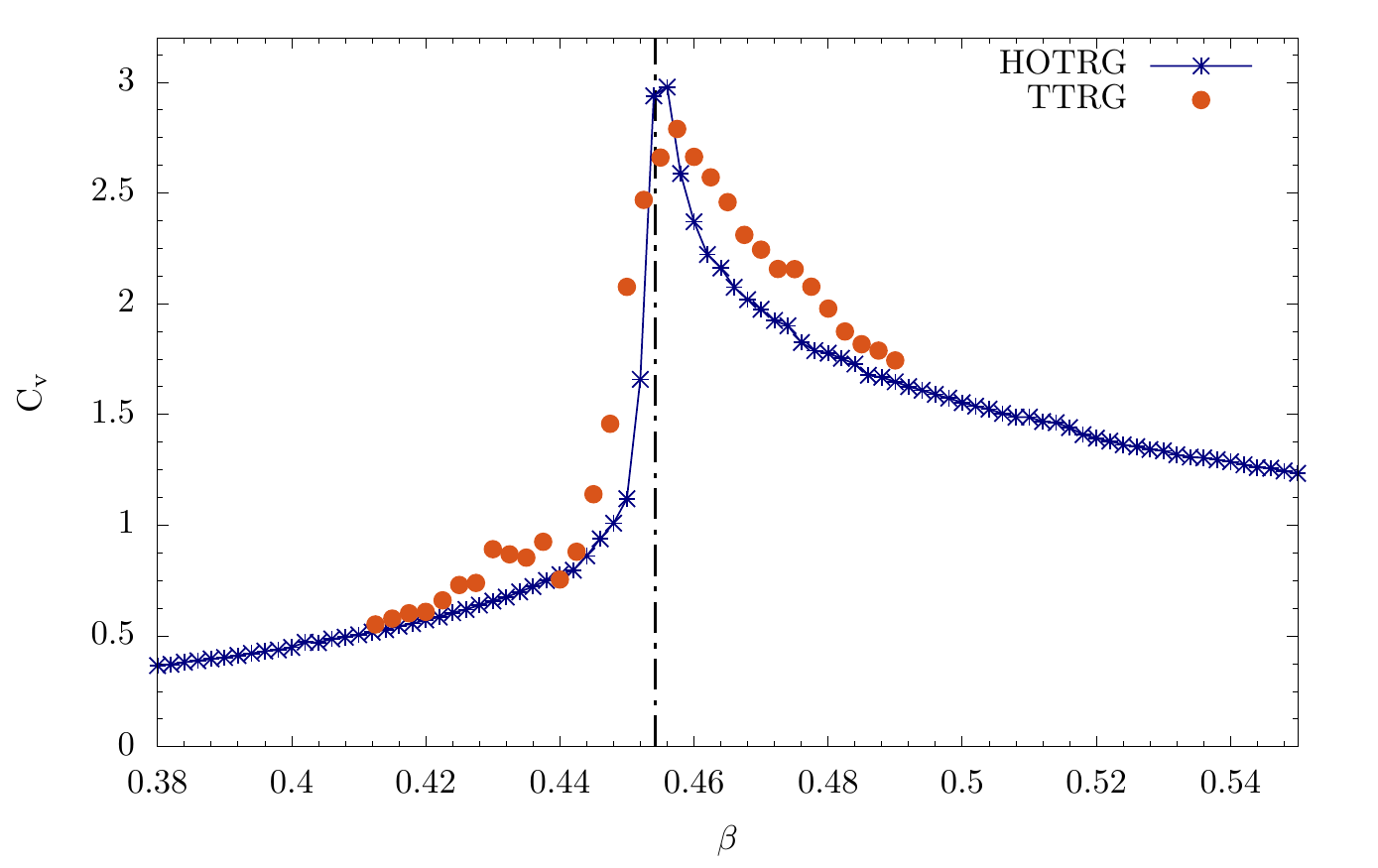}
\includegraphics[width=0.49\textwidth]{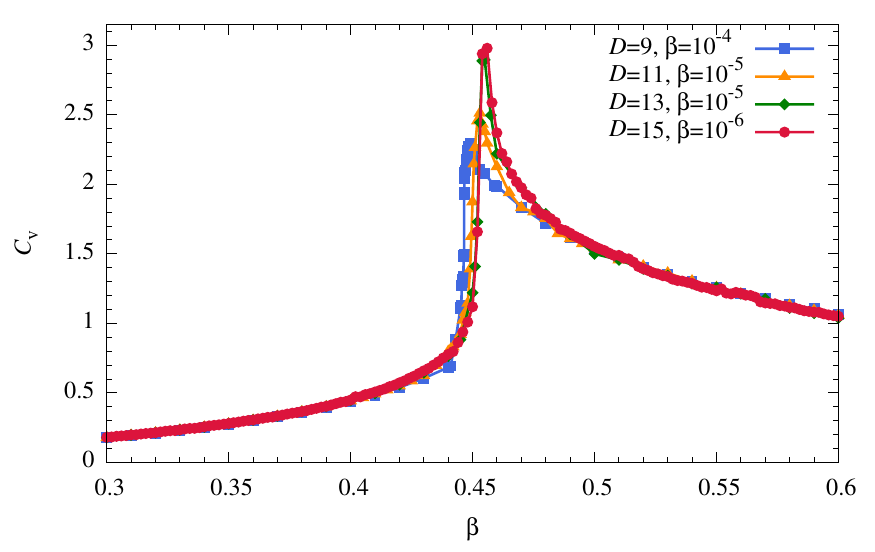}
\caption{Specific heat $C_{\rm v}$ versus coupling $\beta$ on a $32^3$ lattice. Left panel: Comparison of triad TRG with D=72 and a second-order finite difference of $\ln Z$ with step size $\Delta\beta=0.01$, and HOTRG with D=15 using a \textit{stabilized} second-order finite-difference scheme with $\Delta\beta=10^{-6}$. The peak suggests that the critical coupling is between $\beta = 0.45$ and $\beta=0.46$.
For reference, we show the infinite-volume MC result $\beta_{\rm{c}} =  0.454165$ from \cite{Xu:2019mvy} by the black dashed line. Right panel: Convergence of the stabilized finite-difference results for HOTRG with $D=9,11,13,15$.}
\label{fig:specheat}
\end{figure}

Using HOTRG we computed the first and second order derivatives of $\ln Z$ with respect to $\beta$, which yield the internal energy and the specific heat
\begin{align}
C_{\rm v} = \frac{\beta^2}{V} \frac{\partial^2 \ln Z}{\partial \beta^2} .
\end{align}
Computing susceptibilities using HOTRG is quite challenging as second-order finite differences are very sensitive to the systematic errors on $\ln Z$. This is illustrated in the left panel of Fig.~\ref{fig:specheat}  where we show the specific heat computed with second-order finite differences on a $32^3$ lattice using the triad TRG \cite{Kadoh:2019kqk}, which is a state-of-the-art cost-effective variant of HOTRG introducing additional approximations. Even with $D=72$ one has to take $\Delta\beta$ so large to overcome these systematic errors, that a clean investigation of the phase transition region is not possible. 
For this reason we developed a stabilized finite-difference (SFD) method for the standard HOTRG procedure, see Sec.~\ref{sec:improvements} for more details. Although the bond dimension can only be taken up to $D=15$ with HOTRG, the SFD method allows us to reduce $\Delta\beta$ from 0.01 to $10^{-6}$ and the observable is now quite smooth and accurate, as can be seen in the same plot. To verify the results obtained with the SFD method we also investigated the convergence of the transition peak for increasing $D=9,11,13,15$. The right panel of Fig.~\ref{fig:specheat} shows that already for $D=15$ the results are very stable. Although the accuracy of $\beta_c$ is still far from that obtained with Monte Carlo methods, the reproduction of a clean transition peak is an achievement that is far from trivial with tensor-network methods.

\subsection{Number density and Silver Blaze}

One of the main motivations to use tensor-network methods is the possibility to investigate systems with complex actions, which are not accessible to standard importance sampling Monte Carlo methods due to the sign problem. In the presence of a chemical potential, the $O(2)$ model has a strong sign problem, and reweighting, which is a Monte Carlo based algorithm to circumvent the sign problem, notoriously fails.

\begin{figure}
\centering
\includegraphics[width=0.6\textwidth]{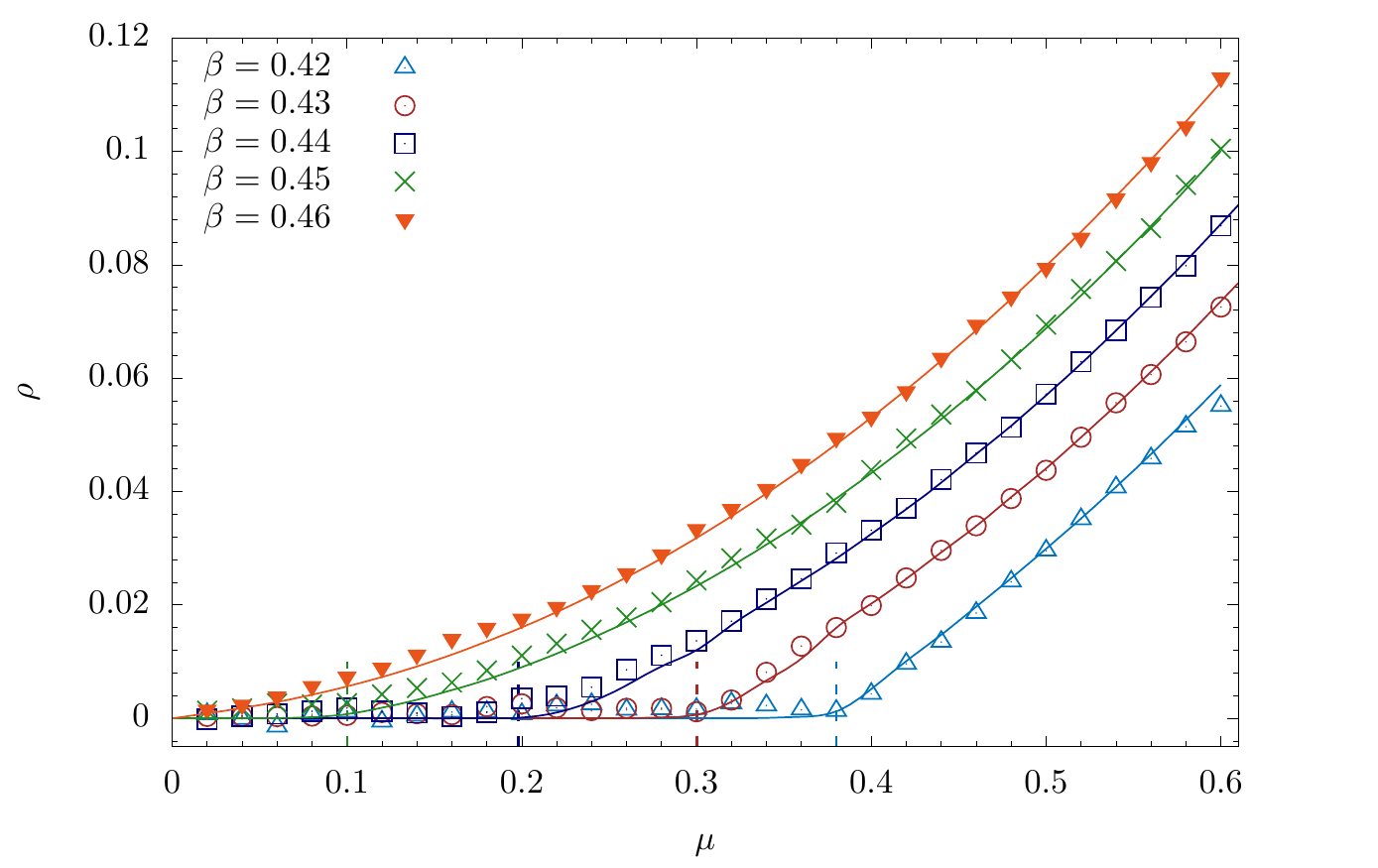}
\caption{Number density $\rho$ versus chemical potential $\mu$ for $T=0$ on a $64^3$ lattice. The data obtained using the triad TRG with $D=50$ (symbols) are compared with those of the worm algorithm (smooth lines). We clearly see the Silver Blaze property of the 3d $O(2)$ model as $\rho$ is zero until $\mu$ reaches a critical value $\mu_c$. }
\label{fig:silverblaze}
\end{figure}

To validate the HOTRG method in this situation we first looked at the number density
\begin{align}
 \rho = \frac{1}{V} \frac{\partial \ln Z}{\partial \mu}
\end{align}
as a function of the chemical potential at zero temperature, i.e., with $N_t=N_s$. As can be seen in Fig.~\ref{fig:silverblaze}, the density remains zero until the chemical potential reaches a critical value $\mu_c$ which corresponds to the mass of the lowest excitation.\footnote{The variation of $\mu_c$ with $\beta$ in Fig.~\ref{fig:silverblaze} is a renormalization effect as the lattice spacing is inversely proportional to the coupling for $\beta \leq \beta_c$, such that $\mu_c= m a(\beta) =\xi^{-1}(\beta)$.}  This property is called the \emph{Silver Blaze} \cite{Cohen:2003kd} and is very hard to reproduce numerically as it is intrinsically coupled to the sign problem in the original spin formulation of the partition function \cite{Aarts:2013bla}. The tensor-network method, on the other hand, does very well to reproduce this property, and the results agree well with those obtained previously with the worm algorithm \cite{Langfeld:2013kno}.

\subsection{Non-zero temperature and chemical potential}

\begin{figure}
\centering
\includegraphics[width=0.6\textwidth]{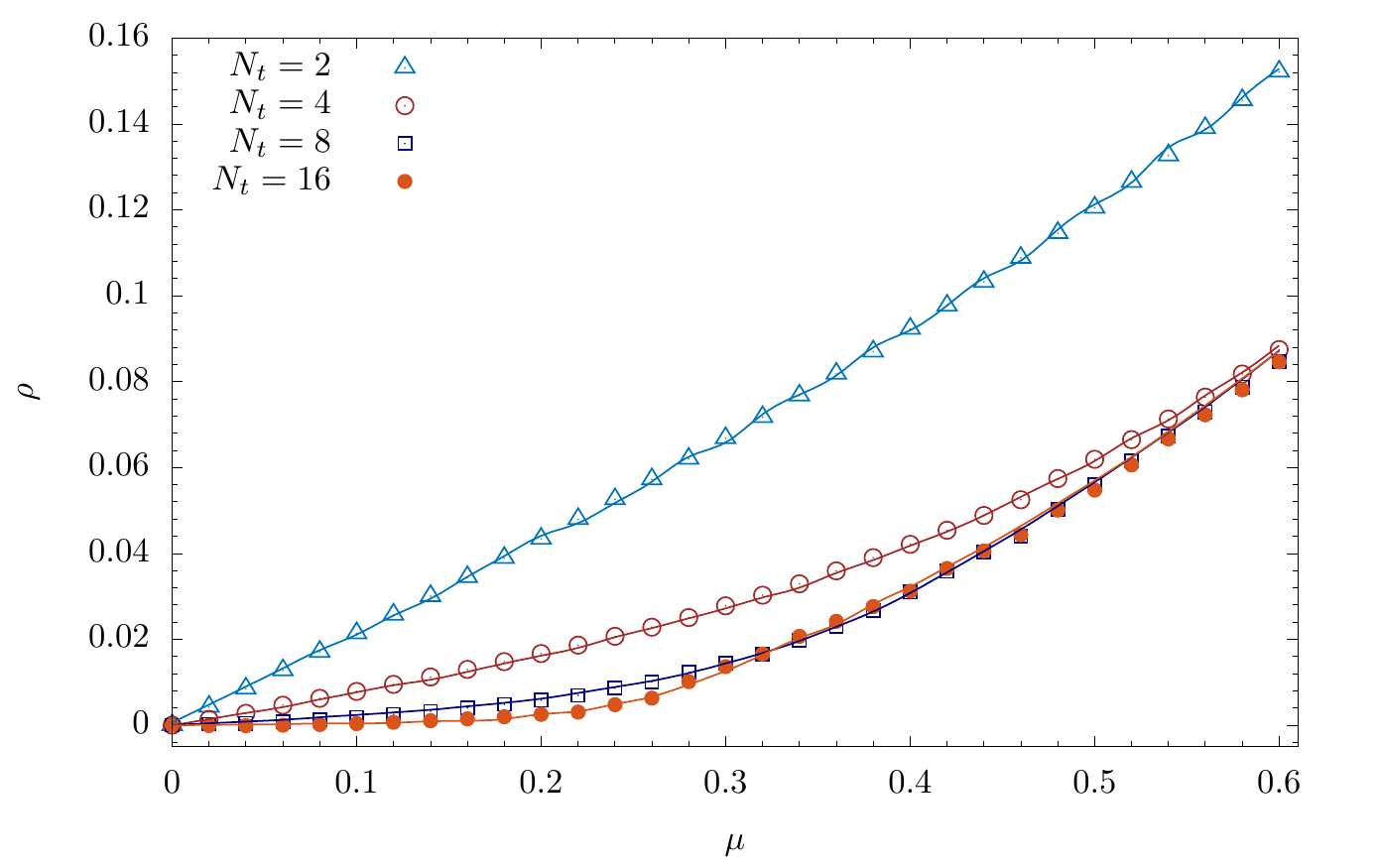}
\caption{Temperature dependence of $\rho$ versus $\mu$ on a $64^2\times N_t$ lattice with $N_t=2,4,8,16$ at $\beta=0.44$. We compare the tensor data obtained using HOTRG with $D=13$, an improved contraction order and stabilized finite differences 
 with data obtained using the worm algorithm (solid lines).}
\label{fig:nonzero-T}
\end{figure}

The tensor-network method can also be used to investigate the model with chemical potential at non-zero temperature, where the time extent of the lattice is reduced such that $N_t<N_s$. Due to the HOTRG blocking procedure it is natural to choose $N_t=2^n$. In Fig.~\ref{fig:nonzero-T} we show the number density on a $64^2\times N_t$ lattice with $N_t=2,4,8,16$ at $\beta=0.44$. For large $N_t$ the results are close to the $T=0$ results of Fig.~\ref{fig:silverblaze}, but as $N_t$ is lowered, higher temperatures are reached and the number density becomes non-zero as soon as $\mu > 0$. This confirms that the Silver Blaze property no longer holds away from zero temperatures. At non-zero temperature, the HOTRG results, which were computed with $D=13$, an improved contraction order and stabilized finite differences, agree well with the results obtained with the worm algorithm. These two improvements to the HOTRG method are briefly discussed below in Sec.~\ref{sec:improvements}.

\subsection{Improvements to standard HOTRG}
\label{sec:improvements}

We briefly describe two improvements to the standard HOTRG method which were used to produce some of the results presented in this paper.

One problem with the standard HOTRG method is that it blocks the lattice by alternating between the different lattice directions. For anisotropic tensors this by no means guarantees to produce a good global truncation error \cite{Bloch:2021mjw}.
We therefore developed an \emph{improved contraction order} where the contraction direction is chosen such that the (approximate) truncation error is minimized in each coarsening step. This procedure only requires a sub-leading additional cost, is especially useful in case of anisotropic tensors and naturally suits the blocking of asymmetric lattices. Although this procedure is local, it has in general a very positive effect on the global truncation error.

The second improvement is related to the computation of observables. The impurity method has a systematic error due to the common truncation frames used for the pure and impure tensors \cite{Bloch:2021mjw}. This error is especially serious in the case of second-order impurities used to compute susceptibilities. On the other hand, the finite-difference method suffers from jumps in $\ln Z$ at nearby parameter values used in the finite-difference formulas, which are caused by level crossings or (almost) degenerate singular values. This generates a very large scatter in the observables, especially in susceptibilities. To resolve this problem we developed a \emph{stabilized finite-difference} method. At each blocking step one first computes the HOSVD factorizations of the blocked tensors for the nearby parameter values used in the finite-difference formula. Then, one analyzes the spaces of singular vectors and performs orthogonal transformations to construct $D$-dimensional subspaces for the nearby parameter values with large singular value content and maximal overlap.

\section{Outlook}

We also made some further improvements that still need some more validation and development. 

In the hierarchical tensor HOTRG (HT-HOTRG) method we use an additional HOSVD factorization of the local tensor to improve the HOTRG efficiency in 3d and 4d \cite{Milde2021}. The basic idea is similar to that of the triads but our approach gathers the backward and forward mode of each direction in a super-mode which is then truncated to a smaller dimension using a HOSVD approximation.

In the iterative Super-Q method we implemented an improved construction of the truncation frames to reduce the total truncation error of the blocked tensor at each coarsening step, whereas the standard HOTRG procedure is  restricted to (partially) reducing the truncation error on the individual modes.

In the future we will extend the stabilized finite differences to factorization schemes like triad-TRG and HT-HOTRG. This is not so straightforward because the additional approximations of these schemes undo the stabilization effects of the SFD method. In another project we will investigate how to use environment information in three- and four-dimensional systems to improve the local truncations in HOTRG, similarly to HOSRG \cite{Xie_2012}.

In other contributions to these proceedings we present tensor-network results of the strong-coupling $U(N)$ model in 3d and 4d \cite{Milde2021} and of an effective $Z_3$ model for QCD at non-zero chemical potential \cite{UnmuthYockey2021}.

In a current project we study the strong-coupling $SU(3)$ gauge theory, which requires the use of Grassmann-HOTRG to handle the baryon-meson system.

\acknowledgments{We thank Raghav G.~Jha for his collaboration on this project.}


\nocite{apsrev42Control}
\bibliographystyle{apsrev4-2.bst}
\bibliography{biblio,revtex-custom} 

\end{document}